\documentclass{llncs}
\usepackage{amsmath}
\usepackage{amssymb}
\usepackage{graphicx}
\usepackage{hyperref}
\hypersetup{colorlinks=false}
\makeatletter
\let\@copyrightspace\relax
\makeatother

\begin{document}
\title{Evaluating community structure in large network with random walks}
\author{Jiankou Li}
\institute{University of Chinese Academy of Sciences, Beijing, P.R.China,\\
\email{lijk@ios.ac.cn}}
\maketitle
\begin{abstract}
Community structure is one of the most important properties
of networks. Most community algorithms are not suitable
for large networks because of their time consuming.
In fact there are lots of networks with millons even
billons of nodes. In such case, most algorithms running
in time $O(n^2logn)$ or even larger are not practical. What we need
are linear or approximately linear time algorithm.
Rising in response to such needs,
we propose a quick methods to evaluate community
structure in networks and then
put forward a local community algorithm with nearly linear time
based on random walks. Using our community evaluating measure, we could
find some difference results from measures used before, i.e., the Newman
Modularity. Our algorithm are effective in
small benchmark networks with small less accuracy than
more complex algorithms but a great of advantage in time
consuming for large networks, especially super large networks.
\end{abstract}
\section{Introduction}
Networks are important tools to study real systems.
Nodes in networks usually organized into 
relative densely groups called communities or clusters.
Community structure have become one of the important
directions. With the computer and internet techniques
developing, networks we could get become larger and larger.  
Take the liveJournal online social network and U.S. patent dataset
for example. LiveJournal is a free on-line community with almost
10 million members \cite{kdd2006group}.
The U.S. patent dataset is maintained by the National Bureau of Economic
Research and includes about 3,923,922 patents \cite{leskovec2005graphs}.
It is reasonable to believe lots of other real networks are larger
and will increase quickly in future. There is a great need for
developing quick community detection algorithms.

To study community structure in large networks, we should
evaluate whether a given network have community structure and how
to find them if there are.
So far, the most accepted measure to evaluate the community 
structure is modularity \cite{clauset2004finding}
\cite{newman2004finding}.
However, modularity has an intrinsic scale
indicting that modules smaller than that scale may not be resolved
\cite{fortunato2007resolution} and finding the partition
with maximal modularity is not a trivial thing.
Local methods which works independent
on the global structure seems more practical in such large datas
\cite{andersen2006local}
\cite{bagrow2008evaluating}
\cite{clauset2005finding}.
One kind of local algorithms divide the whole network into
two parts \cite{bagrow2008evaluating} \cite{clauset2005finding}
\cite{luo2008exploring}, a community $C$ and the set of
nodes with links to $C$, say $B$.
They usually start from a given node $s$ and
then explore $B$ and select one or more nodes to merge into
$C$.  Such operation is repeated until some terminal
condition is satisfied.
Another kind of local algorithms work in some different
ways. Firstly, they calculate a vector around a given node
$s$. This vector includes information that indicates
the tendencies to the $C$ we are to find. Then sort
the vector according the score and a new
vector called a support vector is got.
Finally, we could take a sweep over this support vector
based on some quality and find the community.
One of such algorithms is \cite{andersen2006local} which
has been used by Leskovec to find lots of interesting phenomenon
in real networks
\cite{leskovec2009community}
\cite{leskovec2010empirical}.

As we know, random walks have close relationship with
community. A random walker from a given position will be
'trapped' in a community with high probability.
There are lots of community algorithms proposed inspired by this idea.
A vertices similarity measure and a community similarity measure
are proposed by Latapy and Pons and then communities are get by
a agglomerative procedure \cite{pons2005computing}.
Also based on random walks, Zhou define a distance
between pairs of nodes and use divisive procedure to detect
communities with running time $O(n^3)$
\cite{zhou2003distance}
\cite{zhou2003network}
\cite{zhou2004network}.
Community structure could be related to random walk through
the information theoretic approach where the community detecting
procedure becomes compressing a
description of the probability flow of random walk
\cite{pnas2008maps}. 
Delvenne introduce a quality function indicating the persistence
of clustering over time and unifies the modularity measures
\cite{clauset2004finding}
\cite{newman2004finding}
as well as several definition related with random walk
\cite{pnas2010stability}.
Other community algorithm based on random walk includes
$Markov Cluster Algorithm(MCL)$ with running time
$O(nk^2)$\cite{van2000graph}, methods using signaling process
\cite{hu2008community} and methods of minimizing the matrix
distance\cite{weinan2008optimal}. More detail about these
methods could be found in \cite{fortunato2010community}.
All of these algorithm scale $O(n^2logn)$ or higher,
and are not practical in super large networks with millons
and billons of nodes. In this paper, we give a quick
measure to evaluate the community structure and
a local methods to find communities based on random
walks in nearly linear time, which could be used in 
very large networks.

We arrange the rest of the paper as follows.
In section \ref{sec_random_walk_modularity} we propose
a modularity measure based on random walks.
Then we give a algorithm and test it on
benchmarks in section \ref{section_algorithm}.
The result of some experiments in very large networks are given in
section \ref{sec_application}.
Finally, we give the conclusion in section \ref{section_conclusion}.

\section{Random Walk Modularity}
\label{sec_random_walk_modularity}

As pointed above, the random walk has significant indication
of network structure. We consider the following situation,
a random walk is terminated when it forms a ring and the
corresponding number of steps is called random walk length
(RWL for short). A question of interesting is how long the
expectation of RWL is for a given network? 
We test the relationship between the RWL in ER
random networks \cite{erdos1960evolution}
where every pair of nodes are linked with probability $p$
and the planted $l$-partition model \cite{girvan2002community}
which has been used largely to test a community algorithm's
performance. 

\begin{figure}
\centering
\includegraphics[width=2.5in]{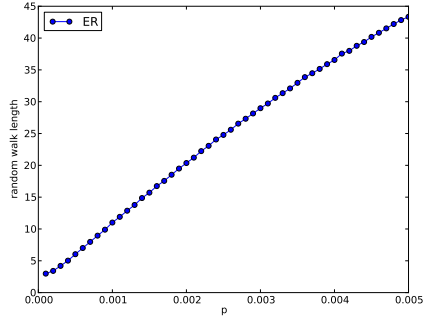}
\caption{The relationship between random walk length and
$p$ in ER random network}
\label{fig_er}
\end{figure}

In the ER random networks where every pair of nodes
are linked with random, the ring are usually formed by two nodes
(the backward walk) as the probability more than two
nodes form a ring is small when the network size tend to infinite.
In every step, the probability a ring is formed by a backward walk
is $\frac{1}{d_v}$. We assume every nodes have the same degree $d$ 
for the sake of discussion.  Let $L_r$ be the expectation of
the random walk length, $q_l$ is the probability that a random
walker ends with l step, we have the following:
\begin{equation}
\begin{split}
L_r = \sum_{l\in n} {q_l}l
   = \frac{1}{d} + 2 * (1 - \frac{1}{d}) * \frac{1}{d}
     + ... + n * (1 - \frac{1}{d})^{n - 1} * \frac{1}{d}\\
   = d - (d + n)(1 - \frac{1}{d})^n
\end{split}
\end{equation}
So when $n\to\infty$ the expectation of RWL is
mainly affected by $d$. When the degree is not constant, the
analysis become complex, but there seems a linear relationship
between the average random walk length and $p$ see figure \ref{fig_er}. 

\begin{figure}
\centering
\includegraphics[width=2.5in]{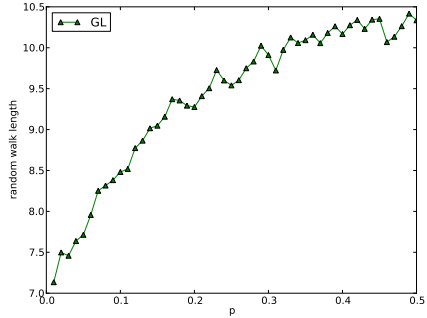}
\caption{Average step length a walker have to pass before
he encounters a ring in $l$-partition model
\cite{girvan2002community}. In this model, every node has a fixed
average degree 16, $p$ is the ratio between the its out degree
and its total degree}
\label{fig_gl}
\end{figure}

On the other hand, the ARL has inverse correlation
with the community structure. The planted $l$-partition
benchmark is used to illustrate the relationship between
random walk length and community structure. This benchmark has been
very popular to test the performance 
of community algorithm since proposed by Condon and Karp
\cite{condon2001algorithms} and a special case of
planted $l$-partition model is given by Newman\cite{girvan2002community}.
In the Newman model, 128 vertices are partitioned into 4 groups with
each group 32 vertices.  Every vertex has $z_{in}$ links in
the same group, and $z_{out}$ links outside of the group. The
average total degree of vertices are fixed to 16.
$p$ is the ratio between $z_{out}$ and average degree of each
vertex. So the community structure could be controlled by $p$.
The average RWL has a close relationship with $p$ see
figure \ref{fig_gl}. This is according with previous idea,
a random walker could be easily trapped in a community.
So the chance that a random walk forms a ring increases in networks
with clear community structure. As $p$ gets larger, the community
structure becomes more and more fuzzy.
As a result, the random walker has more probability to escape
from the 'trap' and RWL increases.

From above analysis, we know RWL is mainly affected by the
degree sequence and the community structure. Inspired by this,
we propose a simple community evaluating measure called Random
Walk Modularity which could be calculated in approximate linear
time. The definition is as follows:

\begin{equation}
Q(G) = 1 - \frac{L(G)}{L(G_r)}
\end{equation}

$L(G)$ means the average random length in graph $G$, and $L(G_r)$ means
the average random length in graph $G_r$ which is the configure model
of $G$ with the same degree sequence of $G$. This measure removes the
influence of degree and reflects community structure
of different networks. 

\begin{figure}
\centering
\includegraphics[width=2.5in]{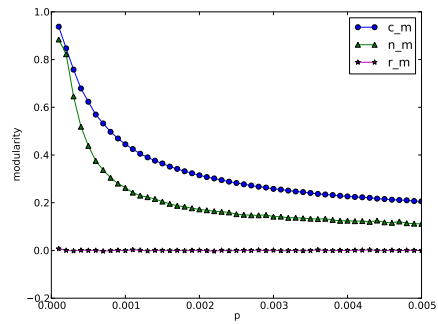}
\caption{Different modularity measures in ER random network,
every pair of vertices are linked with probability p,
c\_m means conductance modularity, n\_m means Newman
modularity, r\_m means random walk modularity}
\label{fig_er_modularity}
\end{figure}

\begin{figure}
\centering
\includegraphics[width=2.5in]{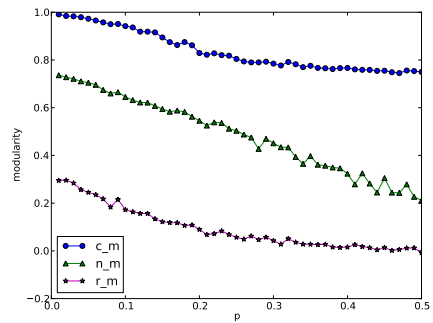}
\caption{Different modularity measures in planted $l$-partition
model, p is the ratio between $z_{out}$ and its average degree,
c\_m means conductance modularity, n\_m means Newman
modularity, r\_m means random walk modularity}
\label{fig_gl_modularity}
\end{figure}

There are some difference
performance in real networks between the Random Walk Modularity
and the modularity \cite{newman2004finding} given
by Newman which we called Newman Modularity in this paper.
Firstly, random network are usually thought to have no
community structure, but when $p$ is small, the Newman
modularity and Conductance Modularity
(see section\ref{section_algorithm}) can be very large.
On the contrary Random Walk Modularity
is not affect by $p$ see figure \ref{fig_er_modularity}, which
is more accord with our intuition. 
Secondly, some deterministic networks, eg. the ring and the lattice, 
have a high Newman Modularity value, but whether they have community 
structure is disputable. In the lattice, every vertex has
the same position, so their community structure even if
they have does not interest us. The Random Walk Modularity
all remove such 
networks by a low value, see table \ref{table_modularity}.
Further more, for networks with clear community structure like the
planted $l$-partition networks, Random Walk Modularity always 
give high value compared with other networks with fuzzy community
structure, see figure \ref{fig_gl_modularity}. 
Finally, the time to calculate Random Walk Modularity is mainly
determined by the average random walk length. Most real networks
have average random length less than 10, and all
of them less than 20 in our experiment, see section \ref{sec_application}.
So Random Walk Modularity can be
calculated in a nearly linear time which indicate that it has
a advantage for large networks. 

\begin{table}
\caption{Different modularity measures
in some deterministic networks. The ring is a one
dimensional lattice, includes 1000 nodes. The tree includes 1000
nodes with all vertices having the same number of children, 2
in our experiments. The lattice have two dimensions, each
dimension has 100 nodes}
\centering
\begin{tabular} {|c|c|c|c|}
\hline
modularity$\backslash$network&ring& tree&lattice\\
\hline
RandomWalk&0.003&0.0004&0.08\\
\hline
Newman&0.94&0.93&0.89\\
\hline
Conductance&0.97&0.93&0.9\\
\hline
\end{tabular}
\label{table_modularity}
\end{table}

\section{Algorithm}
\label{section_algorithm}

In the last section, we propose a quick measure to evaluate
the community structure in large networks.
Starting a random walk from a given node, one could easily be
trapped in a community. The average random
walk length has a inverse correlation with the community
structure. Owing to the community structure,
the average random walk length becomes shorter. 
Our community algorithm are based on this idea.
We perform a series of random walk from a given node $s_0$,
and ends a walk when this walk forms a ring.
At the same time, we assume the tendencies of nodes
to the community of $s_0$ has a positive correlation with
the ring position. Our algorithm is given in the 
following.
\begin{itemize}
\item RandomWalkRing($G$, $s_0$, n)
\begin{enumerate}
\item Set vector $P$ = $\phi$
\item Perform a random walk from $s_0$, record each
      node passed, when encounter a ring ends this walk and
      record its position $l$.
      $\forall v$ in the random walk trail,
      let $P(v) = P(v) + \frac{1}{l * d (v)}$
\item repeat step2 $n$ times
\item Order nodes in $P$ by decreasing value $P(v)$, get a support
      vector S
\item Compute the conductance $\phi(S_i)$ of the
      first $i$ nodes, for i $\leq |S|$
\item find the index $k^*$ at every local optimal of $\phi(S)$,
      return $S_0 = \{v_i | i \leq k^*\}$
\end{enumerate}
\end{itemize}

\begin{figure}
\centering
\includegraphics[width=2.5in]{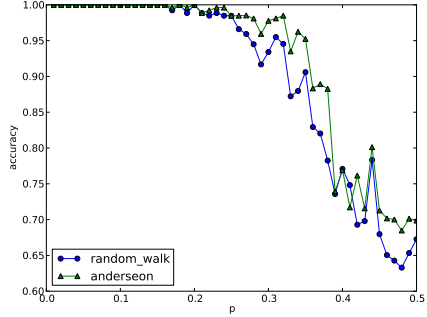}
\caption{The performance of Random Walk Ring and Local Graph
Partitioning algorithm of Andersen}
\label{fig_gl_accuracy}
\end{figure}

In the above algorithm, we need a quality function to
extract the community from the support vector.
We use conductance as the quality function
which has been proposed in
\cite{andersen2006local} and has been used by
Leskovec \cite{kdd2012defining}.
Conductance has been popular to measure community
structure in recent years
\cite{kannan2004clusterings}\cite{leskovec2009community}
\cite{leskovec2010empirical}.
Let the volume $vol(S)$ of a set $S$ be the total degree of
vertices in it, i.e., $vol(S)=\sum_{v\in S}\deg(v)$. 
The conductance $\phi(S)$ of a
set $S$ is defined to be the ratio of the number of edges
$e(S,\bar{S})$ coming out of $S$ with the minimum of the volume of
itself and the volume of its complement $\bar{S}$, i.e.,
$\phi(S)=e(S,\bar{S})/\min\{vol(S),vol(\bar{S})\}$. 
The conductance of the graph is the minimum conductance over all sets
and it is extensively studied in computer science, with applications
to random walks, spectral or flow based graph partitioning, and
combinatorial object constructions.
Intuitively, a set of low conductance (smaller than some
constant $\phi_0$) can be thought of a nice community.
Using this definition, lots of interesting phenomenon have been
found. Leskovec finds most networks seems have a 'core' contains
a constant faction of the nodes with a periphery consisting
of a large number of relatively small 'whiskers'
\cite{leskovec2009community}.

Let $C_1$ and $C_2$ are two communities, we use the following community
similarity measures to evaluate our algorithm, where the planted
$l$-partition model is also used.
\begin{equation}
S(C_{1}, C_{2}) =
\frac{
  |C_{1}\cap C_{2}|
} {
  \sqrt{|C_{1}|*|C_{2}|}
}.
\end{equation}
For each real communities, we find the most similar communities
return by our algorithm, see figure \ref{fig_gl_accuracy} for the
performance of our algorithm and the algorithm of \cite{andersen2006local}.
Our algorithm has similar performance when p is small, and a little
less accuracy when p is larger than 0.25. 
What we emphasize is that our algorithms is very quick in large networks
as pointed before. So such sacrificing of accuracy is inevitable.
Something should be noticed that the accuracy of our algorithm
is affected by n a lot. Generally, the larger n is, the more
the accuracy of our algorithm is. So there is a compromise between
the performance and speed. In our experiment we set n to 1000. 

Conductance is a local definition and could not give a global
knowledge to judge whether a network has good community structure
or not. We give another modularity measure which we called
Conductance Modularity to differ from previous ones.
Let $c$ be a real number between $0$ and $1$, $f$ is the corresponding
fraction of nodes in community with smaller conductance than $c$.
then Conductance Modularity is as follows.
\begin{equation}
C(G) = max_{c\in [0, 1]}\sqrt{(1 - c) * f}
\end{equation}
As we know, usually a smaller conductance indicates the
corresponding community is better. A network with good community
structure should have as many as possible nodes in good communities.
The Conductance Modularity considers both community's quality and 
nodes number, which could give us a intuition whether a network
has community structure or not from the point of conductance.

Figure \ref{fig_gl_modularity} and figure \ref{fig_er_modularity}
are an comparison among three modularity measures.
All of them are sensitive with community structure in
planted $l$-partition model. While Conductance Modularity
and Newman Modularity are affected by $p$ in ER random networks
a lot, Random Walk Modularity is independent on $p$ and seems
more better.
\section{Application}
\label{sec_application}
We perform our algorithm on 34 networks
\footnote{We only consider the corresponding undirected
graphs for all the networks.
Except the citation\_arnerminer network
\cite{tang2008arnetminer} \cite{tang2010combination}
is from \url{http://arnetminer.org/citation} and
the football network is from 
\url{http://www-personal.umich.edu/~mejn/netdata},
all other networks can be found from \url{http://snap.standford.edu}
} in a acceptable time including some very large networks.
Using our algorithms, we could find more than
one communities from each node indicating different
level of communities just as Leskovec do \cite{kdd2012defining}.
In table \ref{table_statistic}, all the results are calculated
from the first local optimal community for the sake of discussion.
As in most case we are more care about the
smallest group includes us, which is always more compact and
has more influence for us although the conductance is not 
the optimal in global.

\begin{table}
\caption{The statistics of network. RM means Random
Walk Modularity, NM means Newman Modularity found by
fast greedy algorithm \cite{clauset2004finding}, CM means
Conductance Modularity, AvgC means the average conductance
of all communities, ARL means average random walk length,
AvgS means the average size of all communities}
\centering
\begin{tabular} {|c|c|c|c|c|c|c|}
\hline
network &RM&NM&CM&AvgC&ARL&AvgS\\
\hline
amazon0302&0.25&0.82&0.79&0.18&5.86&19.95\\
\hline
amazon0312&0.31&0.8&0.73&0.24&8.77&26.29\\
\hline
amazon0505&0.31&0.76&0.73&0.24&8.86&26.42\\
\hline
amazon0601&0.32&0.74&0.73&0.24&8.95&26.96\\
\hline
cit\_arnetminer&0.06&0.65&0.63&0.42&6.63&16.8\\
\hline
cit\_hepph&0.25&0.56&0.55&0.48&18.5&28.7\\
\hline
cit\_hepth&0.27&0.53&0.59&0.42&18.2&31.6\\
\hline
cit\_patents&0.08&0.76&0.59&0.45&8.95&18.6\\
\hline
col\_astroph&0.34&0.51&0.63&0.39&14&22.7\\
\hline
col\_condmat&0.29&0.64&0.74&0.32&6.75&18.3\\
\hline
col\_grqc&0.31&0.79&0.83&0.33&5&17.7\\
\hline
col\_hepph&0.42&0.58&0.71&0.33&11&19.9\\
\hline
col\_hepth&0.19&0.69&0.75&0.33&5.4&16.7\\
\hline
email\_enron&0.18&0.5&0.57&0.47&9.34&49.1\\
\hline
email\_euall&0.01&0.73&0.66&0.46&3.98&409\\
\hline
football&0.17&0.57&0.88&0.14&7.2&20.9\\
\hline
livejournal&0.19&-1&0.53&0.48&15.2&19.3\\
\hline
p2p4&0.007&0.38&0.53&0.6&8.2&11.9\\
\hline
p2p5&0.006&0.4&0.54&0.59&8.1&12.6\\
\hline
p2p6&0.006&0.39&0.54&0.59&8.1&12.5\\
\hline
p2p8&0.015&0.46&0.58&0.54&7.4&12.7\\
\hline
p2p9&0.014&0.46&0.58&0.54&7.3&12.5\\
\hline
p2p24&0.002&0.47&0.62&0.48&5.9&11.1\\
\hline
p2p25&0.005&0.49&0.63&0.47&5.8&11.5\\
\hline
p2p30&0.005&0.5&0.62&0.46&5.8&11.3\\
\hline
p2p31&0.003&0.5&0.63&0.46&5.7&11\\
\hline
roadnet\_ca&0.04&0.99&0.93&0.087&3.67&26.7\\
\hline
roadnet\_pa&0.04&0.99&0.93&0.087&3.68&26.9\\
\hline
roadnet\_tx&0.04&0.99&0.93&0.1&3.63&26.2\\
\hline
vikivote&0.002&-1&0.58&0.5&4.89&496\\
\hline
web\_berkstan&0.54&0.91&0.65&0.32&9&44.4\\
\hline
web\_google&0.39&0.92&0.79&0.17&6.68&30.3\\
\hline
web\_notredame&0.35&0.93&0.76&0.16&5&88.2\\
\hline
web\_stanford&0.47&0.88&0.65&0.35&7.9&40.8\\
\hline
\end{tabular}
\label{table_statistic}
\end{table}

The Random Walk Modularity has a different interpretation
about the community structure compared other measures.
Before discussion, we give the following classification of
network by their corresponding modularity.
Random Walk Modularity in ER random networks is or very near to 0.
So networks with Random Walk Modularity below
0.05 are thought to have no clear community structure, between
0.05 and 0.1 are thought to have weak community structure, and
above 0.1 are thought to have clear community structure.
For Newman Modularity and Conductance Modularity, the boundary
of clear community structure are set to be 0.3 and 0.5 respectively.

From the point of Newman Modularity and Conductance Modularity,
all networks have clear community structure, except livejournal
and vikivote networks whose Newman Modularity could not be
calculated in acceptable time
by the fast greedy algorithm\cite{clauset2004finding}.
The road, web, amazon and some collaboration, citation
and email networks have high value, while others networks have
relative small value. The Newman Modularity and Conductance Modularity
are usually consistent with each other, which means when
one measure give a high score, the other is always give a high
score.

When consider the Random Walk Modularity, the situation
is different. Networks are divided into three classes
as discussed before. The road, p2p, vikivote and email\_euall
networks have no clear community structure, even the road
networks have the highest Newman Modularity and Conductance
Modularity. Citation\_arnetminer and Citation\_ patents networks
has weak community structure. Other networks are thought
to have clear community structure. Such difference
could be explained by figure \ref{fig_er_modularity} and
table \ref{table_modularity}.
Those networks with high Newman Modularity and Conductance
Modularity but small Random Walk Modularity maybe networks
like lattice or with very small average degree whose
community structure are debatable.
In all, Random Walk Modularity are more strict to evaluate
the community structure.

In table \ref{table_statistic}, we also give some other
properties. As a whole, most networks tends to have
small RWL, small conductance and small communities size.
The small communities size maybe influenced by our selection
of the first optimal conductance. The short
AWL is clear, even the most largest networks the liveJournal
online social network and U.S. patent dataset only need about
15 and 9 steps to form a ring. The RWL seems have
a upper bound by the average degree as analysis before.
Owing the influence of community structure, real
RWL is always smaller than that value. The results show Random
Walk Modularity are independent on conductance, RWL and community
size. If a network has high Random Walk Modularity value, we
are more believe it has community structure.

\section{Conclusion}
\label{section_conclusion}
In this paper, we propose a method to evaluate community structure
and a local community algorithm based on random walks with
approximately linear running time.
Our experiments show the average random walk length are affected
by two factors, the average degree of the graph and community
structure. Average random walk length are very short
in real networks, which is either caused by networks' sparseness
or community structure or both. Such short average random
walk guarantees Random Walk Modularity could be calculated
in near linear time. We also give a modularity measure
from the conductance view, which
gives us a profile about a large networks.
Usually the Conductance Modularity and Newman Modularity are
consistent in our experiment, while Random Walk Modularity could
give a different judge.
Random Walk Modularity has advantageous both in evaluating
the community structure
and speed. Networks with high Random Walk Modularity are
more believable to have good community structure, while
Newman Modularity and Conductance Modularity could also give
some ER random network high value. So Random Walk Modularity
should be used when we cared about the
network community structure without the debatable community.

The running time of random ring algorithm is mainly influenced
random walk length and the random walk number.
The former are influence by average degree and community structure
and is usually small, less than 20 in all network in our
experiments. N could be set by user where both accuracy and speed
should be considered. 
Our results show, with some little accuracy sacrifice we could
improve the algorithm's speed a lot. The random ring algorithm
could be used on very large networks with millons or billons
of nodes. 

In the future, we will study the evolution of community structure
and explain why networks form different structures. Methods proposed
in this paper could help disclosed the large network structure a lot.
\bibliographystyle{plain}
\bibliography{mybib}
\end{document}